\documentclass[a4paper]{jpconf}
\usepackage{graphicx}
\def\la{\mathrel{\mathchoice 
{\vcenter{\offinterlineskip\halign{\hfil$\displaystyle##$\hfil\cr<\cr\sim\cr}}}
{\vcenter{\offinterlineskip\halign{\hfil$\textstyle##$\hfil\cr<\cr\sim\cr}}}
{\vcenter{\offinterlineskip\halign{\hfil$\scriptstyle##$\hfil\cr<\cr\sim\cr}}}
{\vcenter{\offinterlineskip\halign{\hfil$\scriptscriptstyle##$\hfil\cr <\cr\sim\cr}}}}}

\def\ga{\mathrel{\mathchoice 
{\vcenter{\offinterlineskip\halign{\hfil$\displaystyle##$\hfil\cr>\cr\sim\cr}}}
{\vcenter{\offinterlineskip\halign{\hfil$\textstyle##$\hfil\cr>\cr\sim\cr}}}
{\vcenter{\offinterlineskip\halign{\hfil$\scriptstyle##$\hfil\cr>\cr\sim\cr}}}
{\vcenter{\offinterlineskip\halign{\hfil$\scriptscriptstyle##$\hfil\cr>\cr\sim\cr}}}}}
\begin{document}
\title{First year results of the High Altitude Water Cherenkov observatory}
\author{Alberto Carrami\~nana - for the HAWC Collaboration}
\address{Instituto Nacional de Astrof\'{\i}sica, \'Optica y Electr\'onica,\\ 
Luis Enrique Erro 1, Tonantzintla, Puebla 72840, M\'exico}
\ead{alberto@inaoep.mx}

\begin{abstract}
The High Altitude Water Cherenkov (HAWC) $\gamma$-ray observatory is a wide field of view (1.8~Sr) and high duty cycle ($>95\%$ up-time) detector of unique capabilities for the study of TeV gamma-ray sources. Installed at an altitude of 4100m in the Northern slope of Volc\'an Sierra Negra, Puebla, by a collaboration of about thirty institutions of Mexico and the United States, HAWC has been in full  operations since March 2015, surveying 2/3 of the sky every sidereal day, monitoring active galaxies and mapping sources in the Galactic Plane to a detection level of 1~Crab per day. This contribution summarizes the main results of the first year of observations of the HAWC $\gamma$-ray observatory.
\end{abstract}

\section{High energy $\gamma$ ray astronomy}
The advent of new spectral windows for the study of the Universe during the second half of the XX$^{\rm th}$ Century brought a remarkable expansion of our knowledge of the cosmos. Today, astronomy is performed across more than twenty decades of the electromagnetic spectrum, from tens of neV, the photon energy of 100 MHz radio-waves, to at least 100 TeV.  Ranging from the MeV to the PeV, gamma rays represent the most extreme form of electromagnetic radiation. Its upper boundary may be even further in energy, having been sought up to the $10^{19}\,\rm eV$ range with the Pierre Auger observatory~\cite{auger-limit}. From the observational point of view, $\gamma$ rays are highly non-thermal radiation associated to astrophysical particle accelerators that produce cosmic rays. And, in the same way as high-energy $\gamma$-ray astronomy is complemented at other wavelengths, it complements searches in the dawn of the multi-messenger era.

Astronomical observations of $\gamma$ rays are performed with space and ground-based instruments. There are three types of detectors with their own particular characteristics, as illustrated in the diagram~\ref{gamma-tel}. Their differences in energy coverage, field of view and sensitivity makes them highly complementary: in the same manner as space-borne pair production telescopes, like {\em Fermi}-LAT or {\em AGILE}, perform continuous observations of large portions of the sky in the GeV range, HAWC permanently surveys the sky transiting above it in the TeV range; both types of instruments provide useful information to atmospheric Cherenkov telescopes, which are highly sensitive but require to be pointed at specific parts of the sky, which are inaccessible part of the year.

\noindent
\setlength{\unitlength}{1cm}
\begin{picture}(16,12)
\put(1.5,6.5){\framebox(7,4.5)[t]{}}
\put(1.5,10.4){\makebox(7,0.5)[t]{\underline{\bf Pair production telescopes}}}
\put(1.5,9.8){\makebox(7,0.5)[t]{0.1 - 100 GeV}}
\put(1.5,9.3){\makebox(7,0.5)[t]{Small effective area}}
\put(1.5,8.8){\makebox(7,0.5)[t]{Background free}}
\put(1.5,8.3){\makebox(7,0.5)[t]{Large FoV and high duty cycle}}
\put(1.5,7.8){\makebox(7,0.5)[t]{\dotfill}}
\put(1.5,7.5){\makebox(7,0.5)[t]{All sky surveys \& monitoring}}
\put(1.5,7.0){\makebox(7,0.5)[t]{Transients (AGN, GRB)}}
\put(1.5,6.5){\makebox(7,0.5)[t]{Extended diffuse emission}}
\put(8.5,6.5){\framebox(7,4.5)[t]{}}
\put(8.5,10.4){\makebox(7,0.5)[t]{\underline{\bf Air shower arrays}}}
\put(8.5,9.8){\makebox(7,0.5)[t]{0.1 - 100 TeV}}
\put(8.5,9.3){\makebox(7,0.5)[t]{Large effective area}}
\put(8.5,8.8){\makebox(7,0.5)[t]{Good noise rejection}}
\put(8.5,8.3){\makebox(7,0.5)[t]{Large FoV and high duty cycle}}
\put(8.5,7.8){\makebox(7,0.5)[t]{\dotfill}}
\put(8.5,7.5){\makebox(7,0.5)[t]{Partial sky surveys \& monitoring}}
\put(8.5,7.0){\makebox(7,0.5)[t]{Transients (AGN, GRB)}}
\put(8.5,6.5){\makebox(7,0.5)[t]{Extended diffuse emission}}
\put(8.5,2.0){\framebox(7,4.5)[t]{}}
\put(8.5,5.9){\makebox(7,0.5)[t]{\underline{\bf Atmospheric Cherenkov telescopes}}}
\put(8.5,5.3){\makebox(7,0.5)[t]{30 GeV - 30 TeV}}
\put(8.5,4.8){\makebox(7,0.5)[t]{Large effective area}}
\put(8.5,4.3){\makebox(7,0.5)[t]{Excellent noise rejection}}
\put(8.5,3.8){\makebox(7,0.5)[t]{Small FoV and low duty cycle}}
\put(8.5,3.3){\makebox(7,0.5)[t]{\dotfill}}
\put(8.5,3.0){\makebox(7,0.5)[t]{Detailed studies of known sources}}
\put(8.5,2.5){\makebox(7,0.5)[t]{Deep surveys of limited regions}}
\put(8.5,2.0){\makebox(7,0.5)[t]{High resolution spectra}}
\put(1.5,2.0){\framebox(7,4.5){}}
\thicklines \put(2.5,2.5){\vector(1,0){5}} \put(7.7,2.4){\mbox{\bf E}}
                 \put(2.5,2.5){\vector(0,1){3}} \put(2.2,5.7){\mbox{\bf FoV}}
                 \put(0.7,3.0){\mbox{Deg}}  \put(0.7,8.5){\mbox{Sr}}
                 \put(5.0,1.6){\mbox{GeV}} \put(12.0,1.6){\mbox{TeV}} 
\put(1.5,0.5){\makebox(14,1){Diagram 1: three types of astronomical $\gamma$ ray telescopes separated in terms}}
\put(1.5,0.0){\makebox(14,1){ of their energy coverage ({\em horizontally}) and field of view (FoV) ({\em vertically}).\label{gamma-tel}}}
\end{picture}

The Large Aperture Telescope (LAT) on board of the {\em Fermi $\gamma$-Ray Space Telescope} has been one of the main driving forces in high energy astrophysics in the last years. Since its 2008 launch, {\em Fermi}-LAT has obtained a deeper and deeper exposure of the whole sky, leading to the finding of about 3000 sources of $\gamma$ rays listed in various catalogs. In particular the 3FGL catalog is the deepest all-sky survey in the 0.1-300 GeV range performed to date, while the 1FHL catalog contains sources detected using photons of energies $>10~\rm GeV$~\cite{3fgl,1fhl}. Of particular relevance for high energy studies is the 2FHL catalog of sources detected in the $50~\rm GeV - 2~TeV$ range, which offers the possibility of complementary studies with atmospheric Cherenkov telescopes and air shower arrays, in particular HAWC~\cite{2fhl}. 

Atmospheric Cherenkov telescopes (ACTs) provide unique detailed information about very high-energy $\gamma$-ray sources~\cite{tev-astro2008}. They employ the atmosphere as the upper part of the detector, taking advantage of particle cascades triggered by the interaction of cosmic and $\gamma$ rays with atmospheric nuclei. ACTs sample the Cherenkov light produced by the secondary particles of the cascade as they travel down the atmosphere faster than the speed of light in the air. As the emission cone of these particles is rather narrow, about $1.4^\circ$, these telescopes have to be pointed to individual objects for their study. Furthermore, the emitted Cherenkov light is faint and can only be detected during night-time, on clear dark nights. To their advantage, ACTs are much more sensitive than air shower arrays: they can detect in a few minutes point sources that require more than one transit with HAWC. In the last couple of decades ACTs have found tens of TeV sources through pointed observations, mostly active galactic nuclei, in particular BL Lacs \& flat spectrum radio quasars, and also starburst galaxies and Galactic sources, as listed in the TeVCat~\cite{tevcat}. An important input to TeVCat is the Galactic Plane survey performed by the HESS collaboration. This is a very deep survey in the TeV band covering Galactic latitudes $|b|<3.5^{\circ}$, and longitudes $250^{\circ} < \ell < 360^{\circ}$ and $\ell < 65^\circ$~\cite{gplane-hess}. 
As a consequence of the large number of Galactic sources found in the HESS Galactic Plane survey, the distribution of source types in TeVCat  departs from the one derived from the 2FHL, highlighting the bias of (deep) pointed observations and highlighting the relevance of TeV wide field of view and high duty cycle observatories. 

Extensive air shower (EAS) arrays can perform continuous survey and monitoring observations complementary to the deeper pointed ACT exposures. EAS arrays have been employed as cosmic-ray detectors for decades; the cosmic-ray experiment installed at Haverah Park in the 1960s is one of the earliest examples of water Cherenkov detectors~\cite{haverah-park}. EAS with the capability of separating photons and hadrons have been developed in United States (MILAGRO), Mexico (HAWC) and China (Tibet AS-$\gamma$ and ARGO).
In particular MILAGRO proved the feasibility of astronomical observations with the water Cherenkov technique by performing a successful multi-steradian survey of the $\gamma$-ray TeV sky. It detected photons of median energy 40 TeV from Galactic objects, in particular extended diffuse sources associated to GeV pulsars. At the Galactic anti-center MILAGRO detected the Crab Nebula and an extended source at the location of Geminga; the first Galactic quadrant showed prominent emission at the Cygnus region and the previously unreported source MGRO~J1908+06. The BL~Lac object Mrk~421 was also detected and sampled in timescales of months during the lifetime of the observatory~\cite{milagro-survey}.

\begin{figure}[t]  \begin{center}
\vspace{-8mm}\includegraphics[width=0.9\hsize]{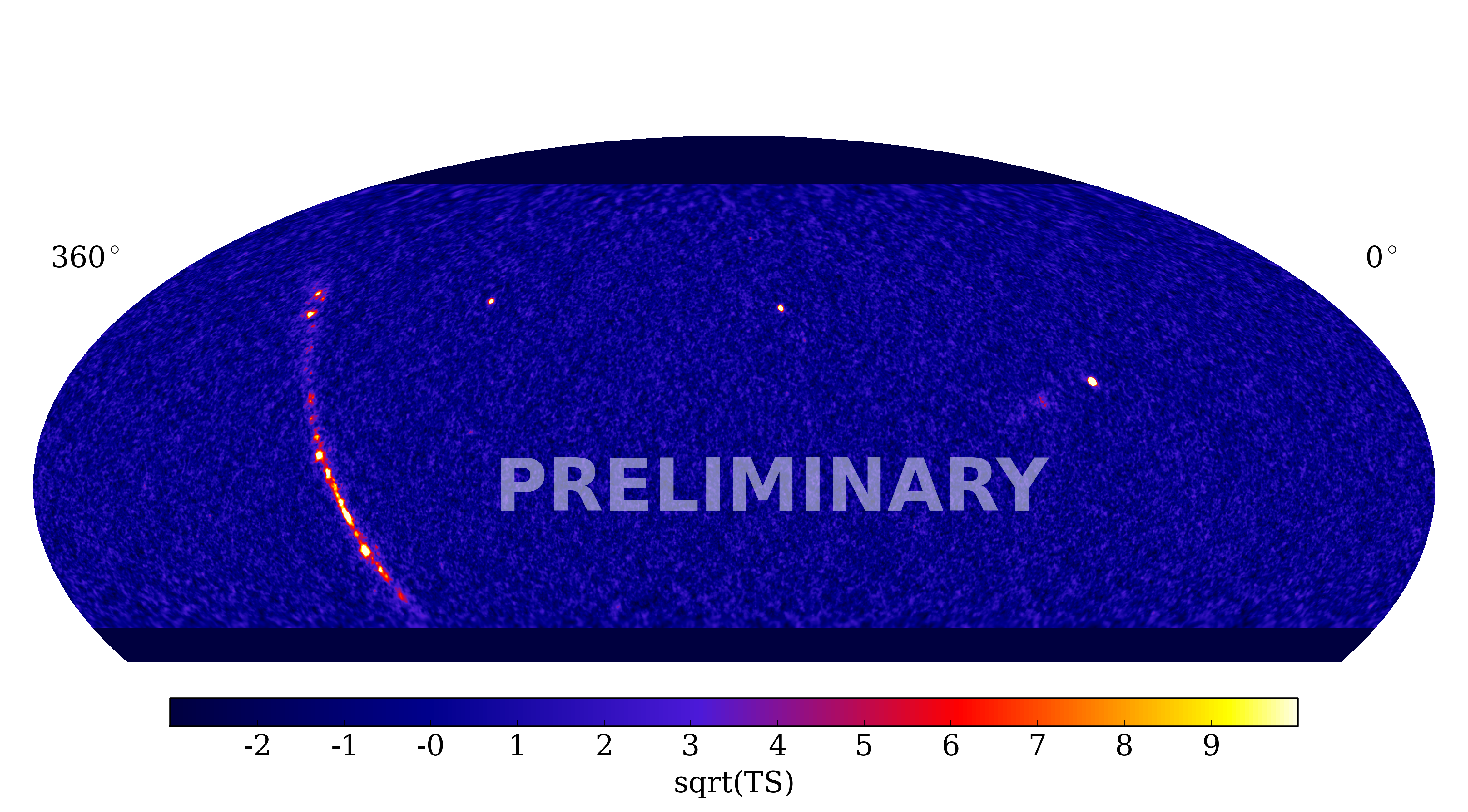}
\end{center}
\vspace{-6mm} \caption{\label{skymap}HAWC skymap of statistical significances above background, in celestial coordinates. The declination coverage is approximately $-30^{\circ}<\delta<+70^{\circ}$. The emission band at the first Galactic quadrant is clearly visible on the left. The three outstanding point sources from left to right, Mrk~501, Mrk~421 and the Crab Nebula. Next to the Crab, with lower significance, is Geminga.} \end{figure}

\section{The High Altitude Water Cherenkov $\gamma$-ray observatory}
The success of MILAGRO prompted for the development of a water Cherenkov observatory of improved design located at higher altitude. The HAWC $\gamma$-ray observatory has been installed at the Northern base of Volc\'an Sierra Negra\footnote{named Tliltepetl, in Nahuatl.}, in the Mexican state of Puebla, by a collaboration of about thirty institutions of Mexico and the United States formed in 2007. The site has an altitude of 4100m and a somewhat equatorial latitude of 19$^\circ$N, which allows to cover the declination band $-26^{\circ}\la \delta \la +64^{\circ}$, or roughly two thirds of the sky for an instantaneous aperture of $45^\circ$. The site selected for HAWC is one kilometer North of the Large Millimeter Telescope Alfonso Serrano (LMT), a 50m diameter antenna for millimeter wave observations installed at 4563m~\cite{gtm}. The base infrastructure is shared between both facilities, as the road, electrical line and optical fiber of the LMT were extended to the HAWC site between 2008 and 2010.
The project funding in place in early 2011 and the HAWC array platform was prepared between May and December 2011. After three years of installation, the complete HAWC array was inaugurated on March 2015 in a ceremony taking place at the 4100m site. 

HAWC has four times the sampling area of MILAGRO and ten times higher muon detection area. The detector is segmented in 300 optically isolated detectors. Each one of these water Cherenkov detectors (WCD) consists of a metallic cylinder of 7.2m diameter and 5m height containing 180,000 liters of treated transparent water, with three 8~inch and one 10~inch up-facing photomultiplier tubes at its bottom. The array occupies an area of 22,500~m$^2$. The central Counting House hosts the data acquisition system, able to cope with a cosmic-ray event rate of 20~kHz. The relative time resolution is of the order of 1~nanosecond, to properly locate events in the sky; the recording of charge deposits allows for photon-hadron discrimination and energy estimation. A laser calibration system is routinely run from the Counting House.

HAWC started early science operations with a partial configuration of 111 detectors in August 2013. The array was upgraded to 250 WCDs in November 2014, four months prior to the start of full operations formalized with the inauguration event on equinox day 2015. The HAWC $\gamma$-ray observatory is currently exploring continuously the sky in the 0.1-100~TeV energy range with an instantaneous field of view of 1.8~Sr and the ability to reject 99.9\% of the registered hadronic events, making it able to detect the Crab Nebula with a signal to noise of $5\sigma$ in a single 6h transit. The increasing understanding of the detector response results in an improved analysis software, currently at its Pass~4 version.

\section{HAWC first year observations}
A skymap of statistical significances, measured by $\sqrt{T_s}$, where $T_s$ is the test statistic resulting of the comparison of a point source hypothesis versus the null hypothesis, was generated for declinations $-30^\circ$ and $+70^\circ$. The HAWC data, acquired between late November 2014 and December 2015, were selected with standard cuts that optimize the response for point sources with a Crab-like spectrum. The map shown in figure~\ref{skymap} shows distinct sources that will be described in detail in the 2HWC catalog, currently in preparation~\cite{2hwc}. They are briefly summarized below.

\subsection{The Galactic anti-center region}
\begin{figure}  
\begin{center}
\includegraphics[width=0.56\hsize,clip,trim=0cm 0cm 0cm 11mm]{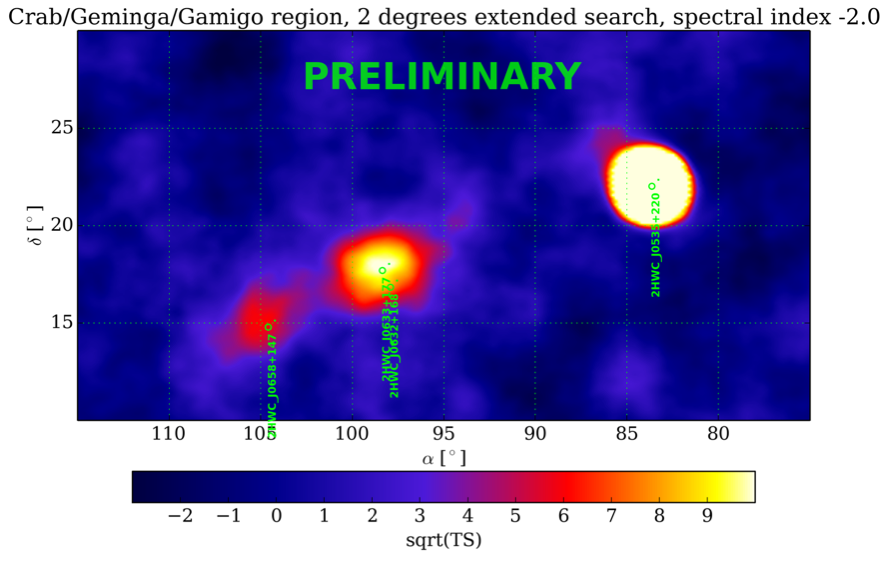} \raisebox{1mm}{\includegraphics[width=0.43\hsize]{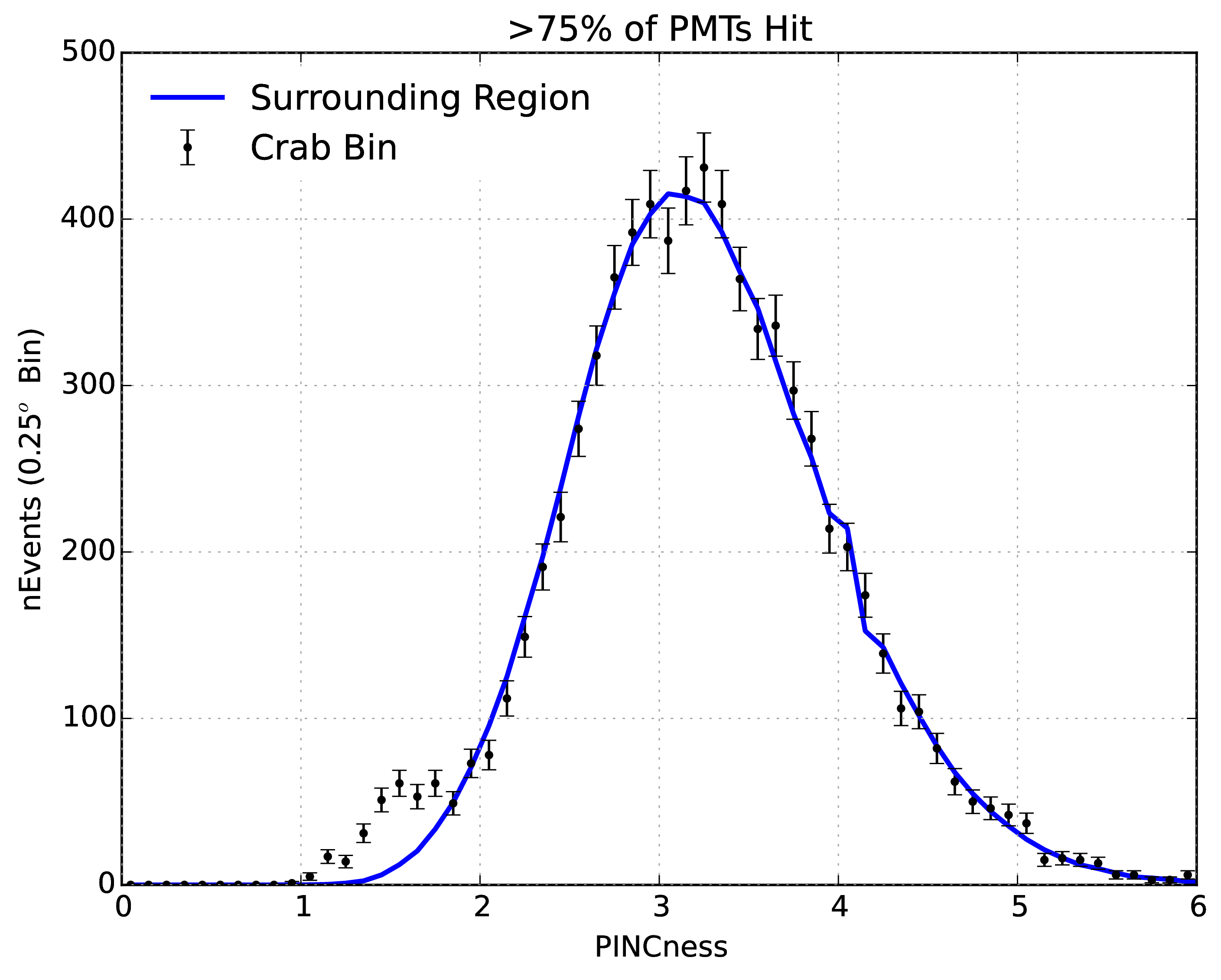}}
\vspace{-4mm}
\caption{\label{anticenter}{\em Left}: Significance map of the Galactic anti-center region. The saturated source on the right corresponds with the Crab Nebula.  The next bright emission region, at $\alpha \sim 98^\circ$, coincides with the position of Geminga, previously reported which MILAGRO. On the left is a likely extended source matching the position of the GeV $\gamma$-ray pulsar PSR~B0656+14. 
{\em Right:} distribution of the PINC parameter for $\gamma$/hadron discrimination on a region excluding the Crab (blue line) and centered on the Crab (black points).} 
\end{center}  
\end{figure}

Three sources have been detected by HAWC in the direction of the Galactic anti-center. The most prominent is the Crab Nebula (M1), a well-known object studied all-through the electromagnetic spectrum. The Crab Nebula and its pulsar are the remnants of SN1054, a supernova recorded by several civilizations over 960~years ago. The proof of their association provided the first hard evidence of the link between supernovae and neutron stars~\cite{m1-sn1054}. The nebula, located at about 2~kpc as estimated through its apparent and physical expansion~\cite{m1-dist}, is powered by the young and powerful 33ms pulsar PSR~J0534+2200.  The {\em Fermi}-LAT spectra of the Crab show the pulsed emission to be dominant between 100~MeV and a few GeV, to be rapidly out-shined by the nebular emission at higher energies~\cite{crab-fermi}. The Crab pulsar has been detected with VERITAS and MAGIC up to $\sim 2~\rm TeV$~\cite{crabpsr-veritas,crabpsr-magic}; however, the pulsed emission may be too faint for a HAWC detection, at least on the short term~\cite{crabpsr-hawc}. The Crab Nebula is a bright source of TeV $\gamma$ rays: it is the brightest source detected by HAWC, with a significance $\ga 80\sigma$ and small enough to be assumed a point source. It serves the TeV community as a calibrator: Crab observations by HAWC allow to compare simulations of its performance with actual data, and to optimize the analysis through improved $\gamma$/hadron cuts~(fig.\ref{anticenter}, right panel). 

One of the most topical current issues is the discovery of flaring activity of the emission of the Crab, reported by {\em AGILE} in 2011~\cite{crabflares-agile}. HAWC has monitored the Crab between November 2014 and February 2016 on a daily basis, obtaining an unprecedented continuous 400-days light curve consistent with a constant flux. We note that, even thought ARGO-YBJ reported an enhancement of the TeV flux of the Crab in 2010, HESS did not find TeV emission associated to the $> 100~\rm MeV$ flaring event of March~2013, and that no MeV flares have been reported since November 2014~\cite{crabflares-argo,crabflares-hess}. HAWC monitoring of the Crab will be relevant to assess its TeV variability and also to explore its spectrum up to 100 TeV~\cite{crab-he-tibet}. 
A dedicated Crab paper is currently in drafting by the HAWC collaboration. 

Geminga is an older 237-ms pulsar, of dynamic age $\sim 300,000$~years, located at a distance of just 130~pc. It is a very bright GeV $\gamma$-ray point source, detected since the 1970-80s with the {\em SAS-II} and {\em COS-B} satellites~\cite{geminga-sas2, geminga-cosb}. {\em ROSAT} and {\em CGRO}-EGRET data showed it contains a radio-silent and $\gamma$-ray loud pulsar~\cite{geminga-pulsar}, becoming the early prototype of this class of objects later unraveled by {\em Fermi}-LAT. The HAWC data are consistent with the MILAGRO observation of an extended emission matching the position of the Geminga pulsar~\cite{milagro-survey}.  

Relatively close to Geminga is the pulsar PSR~B0656+14 (PSRJ0659+1414), detected first in $\gamma$ rays with {\em CGRO}-EGRET and COMPTEL. As in the case of Geminga, this is a relatively old ($\sim 110,000$~years) and nearby pulsar (300~pc), also detected in the radio and optical, but with intense X-ray emission. It is associated with the Monoceros SN remnant~\cite{monoceros-xrays}, itself associated to the EGRET source 2EG~0635+0521; this has been sought, but not detected, in TeV $\gamma$ rays with HEGRA~\cite{monoceros-egret, monoceros-hegra}. HAWC sees an extended emission of a few degrees, difficult to map with ACTs, centered at about J0657+146. The HAWC data of the Geminga and Monoceros regions are currently undergoing detailed studies.

\subsection{Galactic plane}
\begin{figure}  
\begin{center}
\vspace{-2mm}
\includegraphics[width=\hsize]{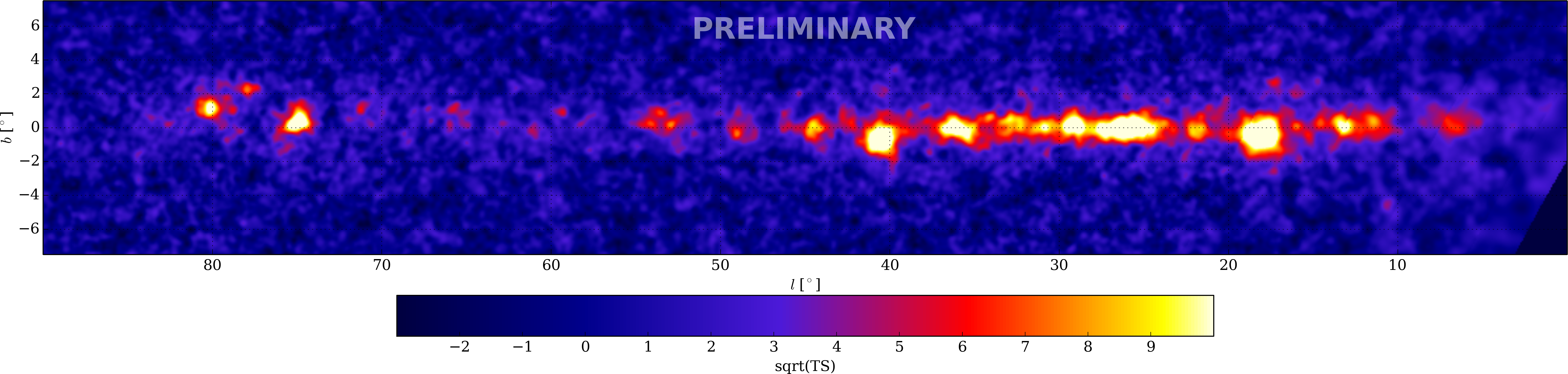}
\caption{\label{gplane}HAWC map of the first Galactic quadrant, displayed in Galactic coordinates. The Cygnus region is at $70^{\circ}\la \ell \la 80^{\circ}$; at $\ell\sim 54^{\circ}$ is the triplet of sources related to SNR~G054.1+00.3 and TeV J1930+188; and several prominent features at lower longitudes, in particular at $40^\circ$ (MGRO~J1908+06) and $18^\circ$ (HESS J1841--055 and J1837--069).}  \end{center}   \end{figure}
HAWC scans every sidereal day close to 70\% of Galactic plane, on longitudes $0^{\circ}\la \ell\la 245^{\circ}$. Aside from the Galactic anti-center region described above, HAWC has observed an extended ridge of $\gamma$-ray emission on the first quadrant of the Galactic Plane, specifically between $5^\circ \la \ell \la 80^\circ$. The Galactic Plane emission was present in the MILAGRO skymap~\cite{milagro-survey}, and is now been studied in more depth with HAWC. Most of the emission is confined to Galactic latitudes $|b|<2^\circ$, indicative of a young population of sources. Deconvolution with a point spread function provides tens of sources, a large fraction of which appear to be extended (fig.~\ref{gplane}). The early HAWC-111 data showed already a good agreement with surveys performed with pointed observations by HESS and VERITAS~\cite{gplane-hess, gplane-hawc111}; this is reinforced by the full-array data. Zones to highlight are: (i) the Cygnus region, which corresponds to a tangent spiral arm, where HAWC sees at least five distinct emissions which can be related to sources like TeV~J2032+4130 (MGRO~J2031+41), VER~J2019+407, and MGRO~J2019+37. These, in turn, correspond well with 3FGL sources; (ii) a complex zone of emission containing the SNR~G054.1+00.3, coincident with TeV~J1930+188, and two previously unreported sources; (iii) a large ridge of $\gamma$ rays emission between $16^{\circ}\la \ell \la 42^{\circ}$, with several regions of intense emission coincident with MGRO~J1908+06, HESS~J1843--033, HESS~J1837--069 and HESS~J1825--137. This emission appears to continue down to the Galactic Center, where the response of HAWC decreases significantly. The Galactic Center itself has not been detected with HAWC, remaining a long term target requiring the full understanding of the high zenith angle response of the detector.

\subsection{Extragalactic sources}
\begin{figure}[b]
\begin{center}
 \includegraphics[width=0.27\hsize]{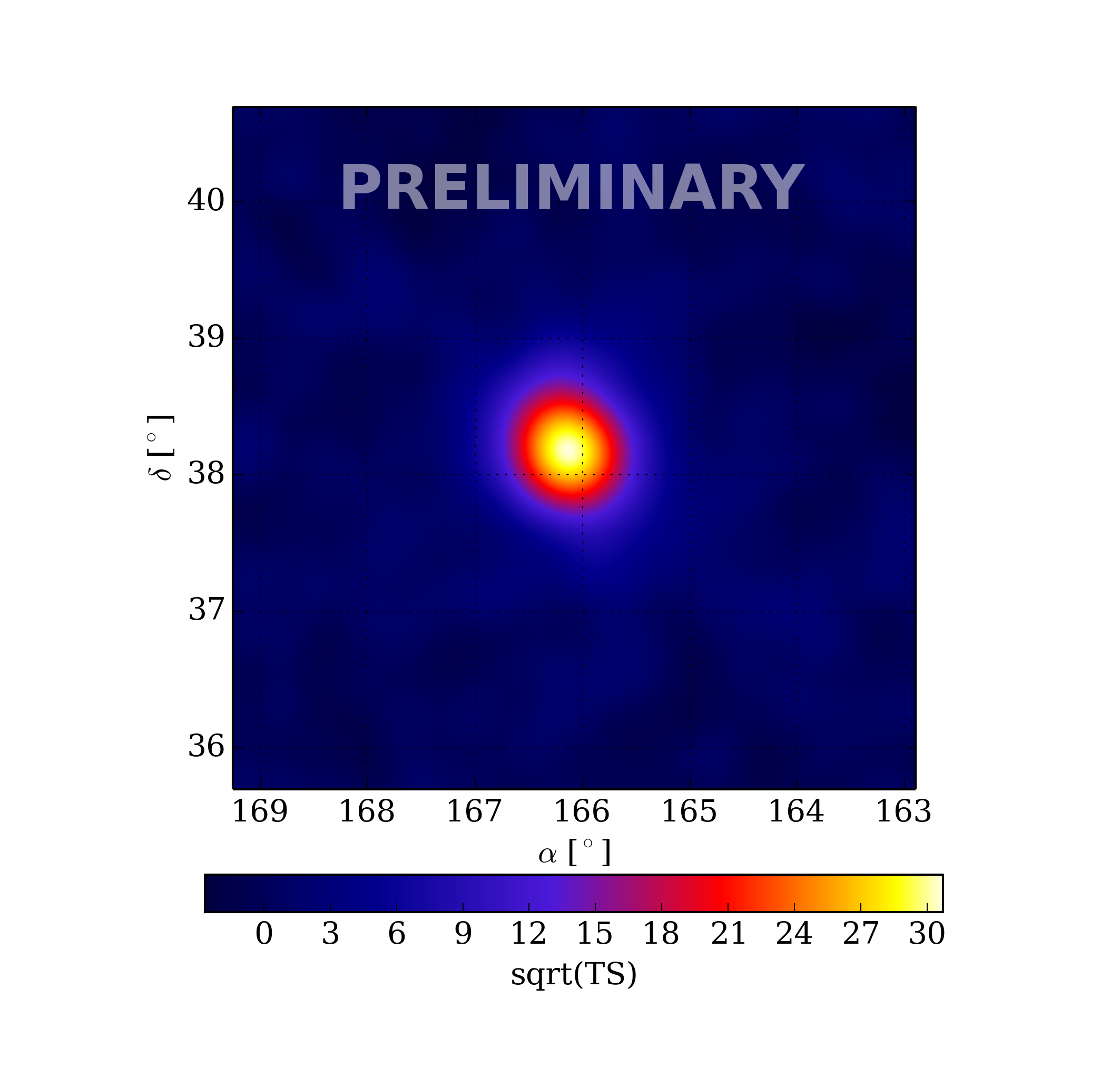} \raisebox{2mm}{\includegraphics[width=0.70\hsize]{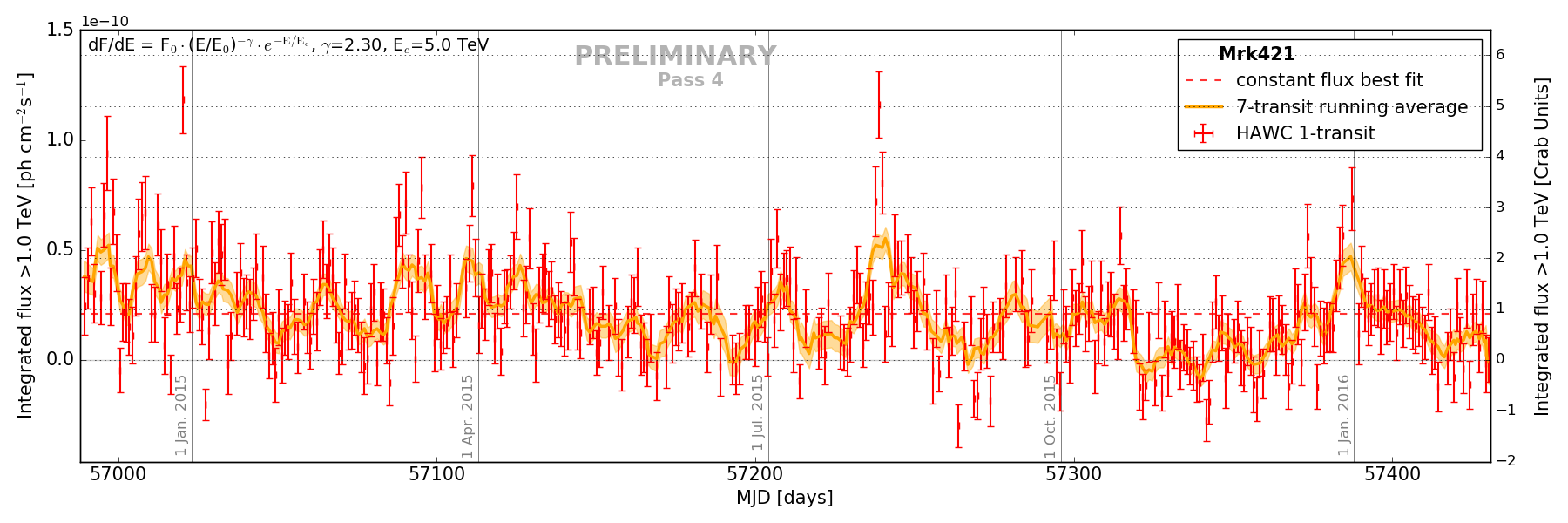}}
\vspace{-4mm}
\caption{\label{421}Detection and one-year light curve of Mrk421. Note the activity around MJD~57240.} \end{center} \end{figure}
\begin{figure}[t]  
\begin{center}
\includegraphics[width=0.27\hsize]{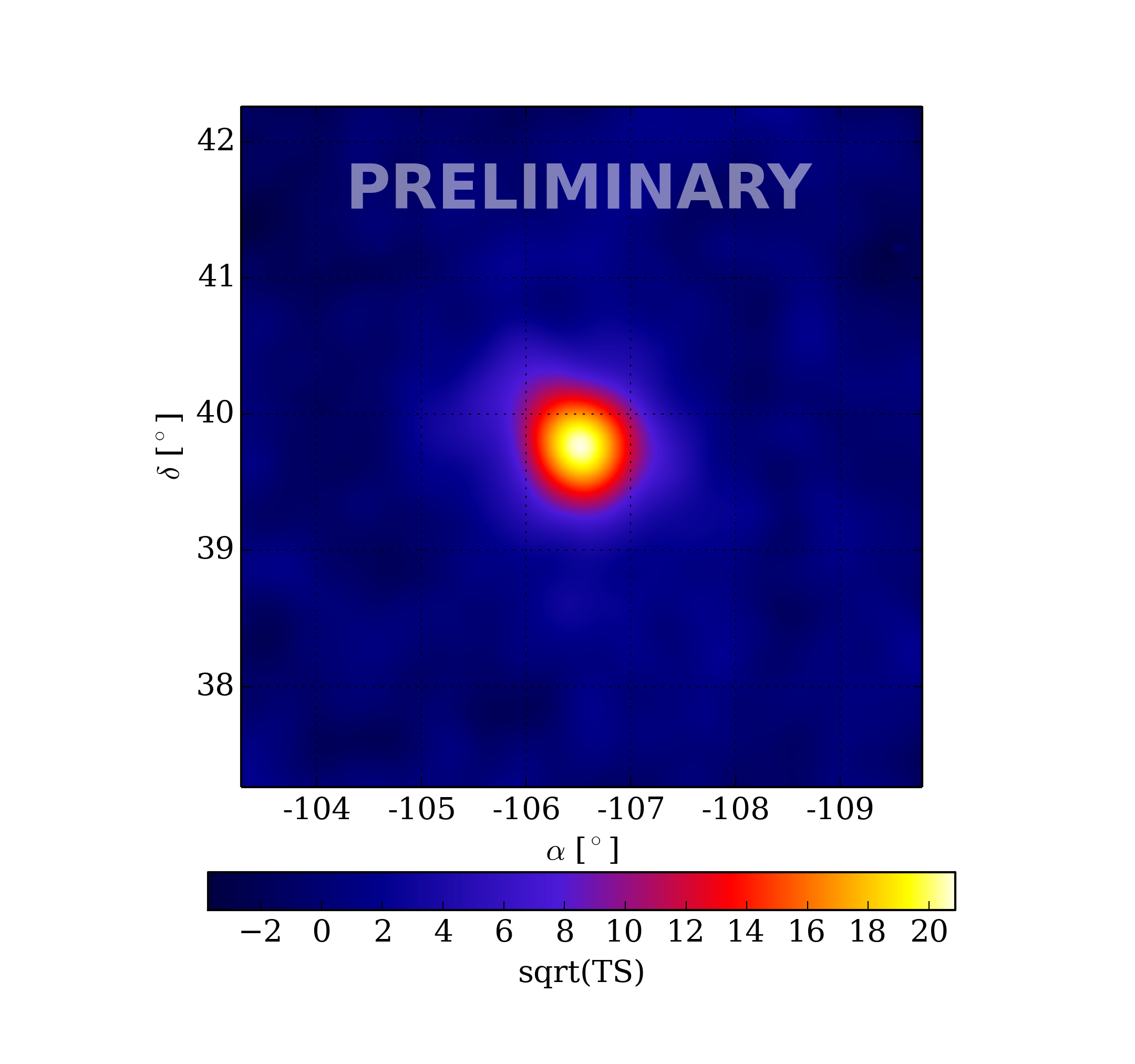} \raisebox{2mm}{\includegraphics[width=0.70\hsize]{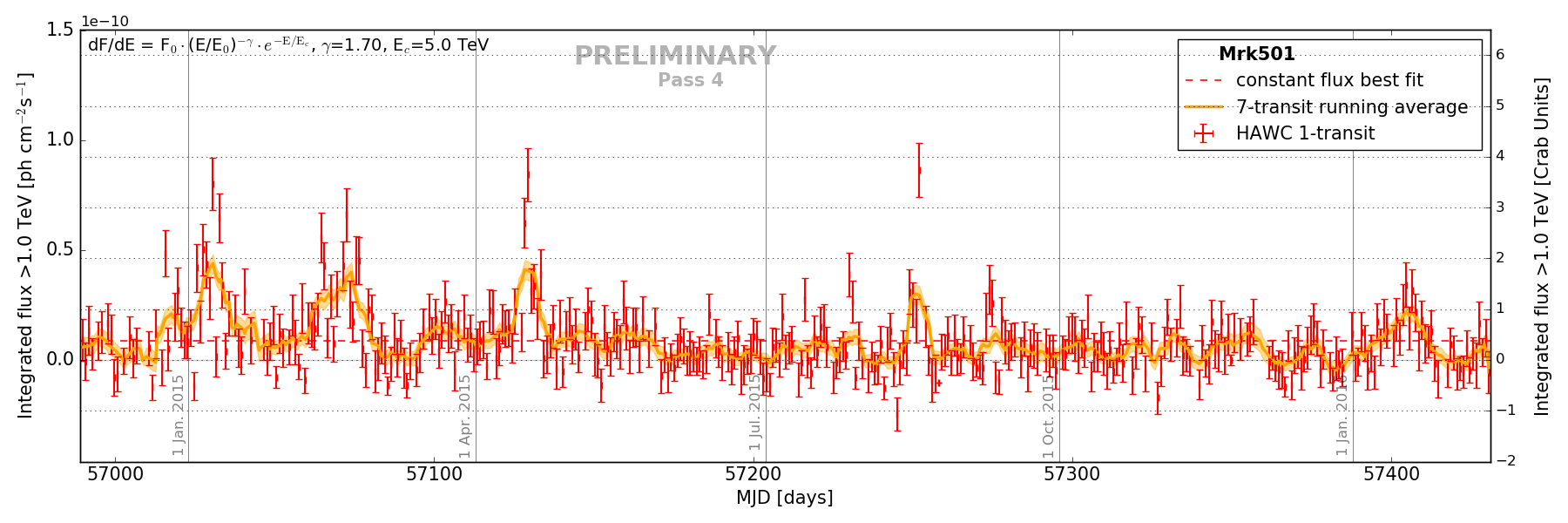}}
\vspace{-4mm} \caption{\label{501}Skymap image and one-year light curve of Mrk~501. A large flare occurring two months after the last data point shown here was reported to the community.} \end{center} \end{figure}

The nearby BL Lac active galaxies Mrk~421 and Mrk~501, previously known to be sources of TeV photons, have been detected by HAWC at high very significant levels, $\sqrt{TS} \simeq 31$ and 22, respectively~\cite{2hwc}. They are bright enough to be frequently detected on weekly timescales, and on single days when flaring. This has allowed to have one-year light curves for each of these Markarian objects with daily sampling~(figs.~\ref{421} and~\ref{501}). Both light curves are inconsistent with a constant flux with $p$-values below $10^{-10}$.

Mrk~421 was discovered as a $\gamma$-ray source by {\em CGRO}-EGRET,  instrument that established active galactic nuclei as the predominant sources of MeV-GeV gamma rays in the sky~\cite{1cgro-cat}. The detections with the Whipple ACT of TeV photons from Mrk~421 {\em only}, the nearest of the first sample of EGRET blazars, represented the first indication of the absorption of very high energy photons by the intervening extragalactic background light (EBL)~\cite{mrk421-tev, ebl-stecker}. Figure~\ref{421} shows the light curve of Mrk~421 from November 2014 to February 2016. This is the first time that Mrk~421 has been sampled continuously for a whole year, as it cannot be observed by ACTs between July and October. When flaring, Mrk~421 can surpass the Crab and become the brightest object in the HAWC sky. The light curve on fig.~\ref{421} shows seven one-day transits with a flux $\geq 4\sigma$ above its average. Episodes of significant enhanced emission can be seen on  timescales of $\sim 10$~days. Note in particular the episode registered around August 2015, {\em i.e.} MJD~57240.

Mrk~501 was discovered as a source of TeV $\gamma$ rays with the Whipple telescope, anticipating by over a decade its finding in GeV energies with {\em Fermi}-LAT \cite{mrk501-tev, mrk501-fermi}. Although not as bright as Mrk~421, this source is detected with over $20\sigma$ and also monitored on daily sidereal basis. Its flux is inconsistency with a constant value, showing significant short timescale activity. Just a few months ago, on the $6^{\rm th}$ of April 2016, HAWC detected Mrk~501 with a flux 2.2 times larger than the Crab Nebula; this was promptly reported to the astronomical community~\cite{mrk501-atel}.

One particularity of these two blazars is their proximity: with a redshift $z = 0.031$, Mrk~421 is at a luminosity distance $d_{L}=134~Mpc$, while Mrk~501 has a slightly larger redshift $z=0.033$ and corresponding to $d_{L}=143~\rm Mpc$. At larger distances the effect of the absorption of high-energy $\gamma$ rays through the pair production process, $\gamma\gamma\to \rm e^{+}e^{-}$, with the EBL is due to become important~\cite{ebl-stecker, ebl-dominguez}. The $\tau=1$ opacity curve can be approximated as $(E/{\rm TeV}) z \simeq 0.15$ in the low TeV band. HAWC is currently performing a search of active galactic nuclei with $z\la 0.3$ to test absorption EBL models. 

\subsection{Fundamental physics}
In addition to the search and study of TeV sources registered in its first year $\gamma$-ray skymap, the HAWC science case considers topics of fundamental relevance for physical sciences, some of which are very briefly highlighted below.

\subsubsection{Cosmic rays --}
With an event rate of about 20,000 cosmic rays per second, HAWC is rapidly building a very large dataset for the study of cosmic particles. Early results confirming the anisotropy in the arrival directions of TeV cosmic rays have already been published~\cite{hawc-cr}. As the cosmic-ray energy range in reach of HAWC is roughly from 1~TeV to 1~PeV, its data can link ground-based PeV observations with the multi-GeV energies reachable with space-borne experiment. The HAWC collaboration is currently working on spectral and composition studies of its extensive cosmic-ray data~\cite{hawc-cr-asp}.

\subsubsection{Gamma-ray bursts --}
The wide field of view and high duty cycle characteristics of HAWC are ideal for searching TeV emission from $\gamma$-ray bursts (GRBs). Even thought these explosive events have been studied in practically all wavebands, it remains to be proven that they do produce TeV photons. {\em Fermi}-LAT detected 35 GRBs in its first three years, some of them with photons in the 50-100~GeV range~\cite{fermi-grbcat}. Even before the start of early science operations in August 2013, HAWC had been seeking to detect GRBs. The exceptional event GRB~130427A, in which a 94~GeV photon was detected by {\em Fermi}-LAT~\cite{fermi-grb130427a}, happened under unfortunate circumstances for HAWC. Then in early phases of installation, the observatory was operating with 29 WCDs and the DAQ system in scaler mode only; the event had a zenith angle of $57^\circ$ and setting. Given these non-optimal conditions, only a very rough upper limit could be set~\cite{hawc-grb130427a}. Just a few months later, better conditions allowed HAWC to start setting constraining upper-limits on later events~\cite{hawc-grb-icrc}. HAWC is constantly searching for GRBs occurring in its field of view: the recent and notorious GRB~160509A, with a multi-GeV photon recorded by LAT, but also with an unfortunately high zenith angle for HAWC, was reported through the GCN system~\cite{hawc-grb-gcn}. 

\subsubsection{Dark matter searches --}
HAWC is well suited to search for $\gamma$ rays produced by the decay of weakly interacting massive particle (WIMPs). The 5-year sensitivity of HAWC for the dark matter emission expected for the Virgo cluster, M31 and the Galactic Center were shown at the ICRC2015~\cite{hawc-darkmatter}. The nearby Andromeda galaxy is probably the best candidate at low WIMP masses, while the Galactic Center may provide the best limits at higher energies, $M_\chi \ga 30 - 100~\rm TeV$, despite its unfavorable location relative to the Sierra Negra site~\cite{hawc-darkmatter}.

\subsubsection{Primordial Black Holes --}
Primordial black holes (PBHs) are remnants of the Big Bang that may have formed during phase transitions related to collapses of cosmic strings or domain walls, in the very early Universe~\cite{pbh-creation}. According to quantum gravity theories, they evaporate through Hawking radiation following the formalism of classical thermodynamics extended to gravitation~\cite{pbh-hawking, pbh-gbursts}. PBHs radiate according to their mass on timescales $\tau(M)\propto M^3$, such that black holes with $M_{BH}\la 5\times 10^{14}\rm g$ should be evaporating now. Within the uncertainties on how the final stages of the evaporation process lead to an explosive event, PBHs may have a distinct signature observable in the GeV/TeV range~\cite{pbh-gamma}. The most constraining upper limits of the volume density of PBHs decaying has been set with MILAGRO and will be superseded by HAWC with its first year data, to be gradually improved as the HAWC database builds up~\cite{pbh-milagro}.

\subsubsection{Neutrino astrophysics --}
The IceCube neutrino observatory is providing a new fresh look at astrophysical phenomena in the high-energy regime. Astrophysical neutrinos are a powerful diagnostic of high-energy hadronic processes, which necessarily produce $\gamma$ rays. HAWC is coordinating efforts with IceCube in the search for enhanced TeV photon emission related to multi-TeV and PeV neutrino events in real-time observations, and also in archival data~\cite{neutrino-atel, neutrino-gcncirc}.

\subsubsection{Gravitational wave events follow-up --}
The spectacular discovery of a gravitational wave emission event observed by LIGO on the 15$^{\rm th}$ of September 2015, interesting in its own right as an astrophysical black hole merger, is a landmark in the study of the Universe~\cite{ligo-gw150915}. 
Since 2014 the HAWC collaboration has worked in coordination with the LIGO/VIRGO collaboration to search for high energy electromagnetic counterparts of gravitational wave events. The GW150915 discovery event occurred towards the celestial South Pole, making unfeasible such a follow-up. The localization by LIGO of the more recent GW151226 event, a coalescence of two black holes of masses 14 and 7~$M_\odot$ respectively~\cite{ligo-gw151226}, intersected the field of view of HAWC at the time of the event. The analysis of these HAWC data is on-going.

\section{HAWC summary}
The HAWC $\gamma$-ray observatory has become a prime survey instrument for high-energy astrophysics. Its first year data have provided a novel view of the TeV sky, with prominent sources in the Galactic anti-center, the first Galactic quadrant and outside our own galaxy. A deeper understanding of these, together with studies in fundamental topics like cosmic rays, GRBs, dark matter searches, PBHs and multi-messenger follow-ups, makes us look forward to further HAWC data. The HAWC collaboration is currently working on upgrading its facilities with a high bandwidth internet optical fiber connection from the Sierra Negra site to the INAOE, and a sparse 100,000~m$^2$ outrigger array to improve the response of the instrument above 10~TeV.

\section*{Acknowledgements}
The High Altitude Water Cherenkov $\gamma$-ray observatory is a collaboration of over thirty institutions in Mexico and the United States, with current participation of Germany, Poland and Costa Rica. HAWC has been possible thanks to the generous support of the National Science Foundation, the US Department of Energy and the Consejo Nacional de Ciencia y Tecnolog\'{\i}a in Mexico.

This research has made use of the SIMBAD database, operated at CDS, Strasburg, France~\cite{simbad}; and of NASA's Astrophysics Data System~\cite{ads}.

\end{document}